\documentclass[11pt,a4paper]{article}
\usepackage{amsmath}
\usepackage{float}
\usepackage{caption2}
\usepackage{graphicx}

\begin{document}

\title{\textbf{\large{Symmetries of the Ricci Tensor of Static Space-times with Maximal Symmetric Transverse Spaces}}}
\vspace{1cm}
\author{M. Akbar\thanks{Electronic mail: akbar@itp.ac.cn}\\
\textit{Institute of Theoretical Physics, Chinese Academy of
Sciences} \\ {P. O. Box 2735, Beijing 100080, China}}
\date{}
\maketitle
\begin{abstract}

Static space times with maximal symmetric transverse spaces are
classified according to their Ricci collineations. These are
investigated for non-degenerate Ricci tensor ($det.(R_{\alpha})
\neq 0$). It turns out that the only collineations admitted by
these spaces can be ten, seven, six or four. Some new metrics
admitting proper Ricci collineations are also investigated.

\vspace{.5cm}\noindent PACS numbers: 04.20.-q, 04.20.Jb
\end{abstract}

\section{Introduction}\label{sec1}
Let \textbf{M} be a four dimensional smooth, connected, hausdorff
manifold admitting a smooth Lorentz metric $g$ of signature (+ - -
-). A Lie derivative along a vector field \textbf{V} is denoted by
$L_{\textbf{V}}$, when component notation is used, a partial
derivative and a covariant derivative with respect to the
Levi-Civita connection $\Gamma$ associated with $g$ are denoted by
a comma and a semi-colon respectively. In general relativity
theory \textbf{M} plays the role of the space time and the
geometrical objects $g$, $\Gamma$ and the curvature tensor on
\textbf{M} derived from $\Gamma$ collectively describe the
gravitational fields. Einstein's equations

\begin{equation}\label{1}
G_{\mu\nu} = R_{\mu\nu} - \frac{1}{2} R g_{\mu\nu} = \kappa
T_{\mu\nu}
\end{equation}
provide physical restrictions on these objects. Where $G_{\mu\nu}$
is the Einstein tensor, $R_{\mu\nu}$ is the Ricci tensor, $R =
g^{\mu\nu}R_{\mu\nu}$ is the Ricci scalar and $T_{\mu\nu}$ is the
energy-momentum tensor. I have assumed here that the cosmological
constant is zero. Using the Bianchi identity, it can easily be
shown that $G^{\mu\nu}_{; \nu} = 0 \Leftrightarrow T^{\mu\nu}_{;
\nu} = 0$. A space time is said to permit collineation if
\begin{equation}\label{2}
L_{\textbf{V}}\phi = \Lambda,
\end{equation}
where $\textbf{V}$ is the collineation vector, $\phi$ is any of
the quantities, $g_{\mu\nu}$, $\Gamma^{\lambda}_{\mu\nu}$
$R_{\mu\nu}$, $R^{\alpha}_{\lambda\mu\nu}$ and geometric objects
derived from these quantities. $\Lambda$ is a tensor with the same
index symmetries as $\phi$. Hence many types of symmetries
structure of the space time are investigated in general theory of
relativity in order to understand the natural relationship between
geometry and matter furnished by the Einstein field equations
\cite{a1}, \cite{a2} . Isometries, Homothetic motions, conformal
Isometries, affine and projective collineations and symmetries of
curvature and related tensors are the examples of these symmetries
\cite{a3} . These symmetries structure of the space times not only
help us to obtain exact solutions of Einstein field equations but
also provide invariant bases for classifying space-times. In
recent years, a large number of exact solutions of Einstein field
equations with different symmetry structures have been found
\cite{a4} and classified space time according to their symmetries
structure \cite{a5}. A space time is said to admit Ricci
collineation vector field \textbf{V} on Lorentzian manifold
\textbf{M} if the Lie derivative of $R_{\mu\nu}$ along \textbf{V}
is zero.
\begin{equation}\label{3}
(C_{\mu\nu}) = L_{V}R_{\mu\nu} = 0 \Longleftrightarrow R_{\mu\nu},
_{\lambda}V^{\lambda} + R_{\mu\lambda}V^{\lambda}, _{\nu} +
R_{\nu\lambda}V^{\lambda}, _{\mu} = 0
\end{equation}
The Ricci collineations are purely geometric in nature but like
matter collineations provide physical information of space-times
through Einstein field equations. In recent years, there has been
much concern in the study of the various symmetries, particularly
in matter and Ricci collineations. Green et al \cite{a6} and Nunez
et al \cite{a7} have considered an example of Ricci collineation
and the family of contracted Ricci collineation symmetries of
Robertson-Walker metric. They have restricted their study to
symmetries generated by the collineation vector field of the
following form, respectively,
$$\textbf{V} = V^{4}(t, r, \theta, \phi) \frac{\partial}{\partial
t}~~~and ~~~~ \textbf{V} = V^{1}(t, r) \frac{\partial}{\partial r}
+ V^{4}(t, r) \frac{\partial}{\partial t}$$ Also, the relationship
with the constants of motion between Ricci collineation and family
of Contracted Ricci collineation has been investigated in
references \cite{a8}, \cite{a9}, \cite{a10},\cite{a11}. Ricci and
matter collineations for static spherically symmetric space times
have been studied recently by various authors \cite{a12},
\cite{a13}, \cite{a14},\cite{a15}. Amir, Bokhari and Qadir has
found the relationship between the Ricci collineations and Killing
vectors for these space times \cite{a12}. Contreras et al
\cite{a16} have investigated Ricci collineations for non-static
spherically symmetric space times and they have confined their
study to the symmetries generated by the vector field of the form,
$$\textbf{V} = V^{t}(t, r)\frac{\partial}{\partial t} + V^{r}(t,
r) \frac{\partial}{\partial r}$$ Qadir, Saifullah, and Ziad
\cite{a17} have considered cylindrically symmetric static space
times and worked out complete classification according to their
Ricci collineations. In addition, Caret et al \cite{a18} has
studied matter collineations as a symmetric property of the
energy-momentum tensor and Hall et al \cite{a19} have studied the
Ricci and matter collineations. Recently Camci et al \cite{a20}
have classified Bianchi types-1 and 111, and Kantowski-Sachs space
times according to their Ricci collineation vector field. Petrov
\cite{a21} has initiated an approach of finding information about
the solutions of Einstein field equations without specifying the
stress-energy tensor, instead of looking only at the space-time
symmetries. This approach does not always provide specific
metrics, or classes of metrics for a given isometry. In fact some
symmetries were given in reference \cite{a22} for which there is
no corresponding metric. Subsequently an approach was developed to
ask for minimal isometry group and then classify completely all
higher symmetry space times. This
method provides complete classification for various space times.\\
The general line element for a static space times with maximal
symmetric transverse spaces can be written as
\begin{equation}\label{4}
ds^{2} = e^{A(r)}dt^{2}-e^{B(r)}dr^{2}-r^{2}(d\theta^{2} +
f^{2}_{k}(\theta)d\phi^{2})
\end{equation}
where $f_{k}(\theta)$ is given by
$$f_{k}(\theta) = sin\theta ~~~~~~~ when ~~k = 1$$

$$~~~~~ = \theta~~~~~~~~~~ when ~~k = 0$$

$$ ~~~~~~~~~~~ = sinh\theta~~~~~~~~~ when ~~k = -1$$
The maximal symmetric transverse spaces are Einstein spaces. As
the metric under consideration is diagonal and the metric
coefficients depend on $r$ only, the non-zero components of Ricci
tensor are
\begin{equation}\label{5}
R_{00} = (\frac{A^{\prime\prime}}{2} -
\frac{A^{\prime}B^{\prime}}{4} + \frac{A^{\prime 2}}{4} +
\frac{A^{\prime}}{r}) e^{A - B}
\end{equation}
\begin{equation}\label{6}
R_{11} = \frac{-A^{\prime\prime}}{2} +
\frac{A^{\prime}B^{\prime}}{4} - \frac{A^{\prime 2}}{4} +
\frac{B^{\prime}}{r}
\end{equation}
\begin{equation}\label{7}
R_{22} = (\frac{rB^{\prime}}{2} - \frac{rA^{\prime}}{2} - 1)
e^{-B} + k
\end{equation}
\begin{equation}\label{8}
R_{33} = f_{k}^{2}(\theta) R_{22}
\end{equation}
Here prime indicates the derivative with respect to $r$. Let
$R_{\alpha\alpha}$ is denoted by $R_{\alpha}$, where the Greek
indices $\alpha$, $\beta$ will be used when it is needed to drop
the summation convention. Further $R_{\alpha}$ are arbitrary
function of $r$, for $\alpha = 0,~1~~2$ and $R_{3} =
R_{2}f_{k}^{2}(\theta)$. Using the components of Ricci tensor for
equation (4) in Ricci collineation equations (3) I get the
following independent equations
\begin{equation}\label{9}
(C_{\alpha\beta}) = R_{\alpha}V^{\alpha}_{,\beta} +
R_{\beta}V^{\beta}_{,\alpha} = 0
\end{equation}
\begin{equation}\label{10}
(C_{00}) = R^{\prime}_{0}V^{1} + 2R_{0}V^{0}_{,0} = 0
\end{equation}
\begin{equation}\label{11}
(C_{11}) = R^{\prime}_{1} V^{1} + 2R_{1}V^{1}_{,1} = 0
\end{equation}
\begin{equation}\label{12}
(C_{22}) = R^{\prime}_{2}V^{1} + 2R_{2}V^{2}_{,2} =0
\end{equation}
\begin{equation}\label{13}
(C_{33}) = R^{\prime}_{2}V^{1} + 2R_{2}[(\frac{f_{k~,2}}{f_{k}})
V^{2} + V^{3}_{,3}] = 0
\end{equation}
Equations (9) give six equations and equations (10-13) are four
partial differential equations, these constitute together ten
first order, non-linear coupled partial differential equations
involving four components of the arbitrary Ricci collineation
vector $\textbf{V} = (V^{0}, V^{1}, V^{2}, V^{3})$, four
components of Ricci tensor and their partial derivatives. The
components $V^{\mu}$ with $\mu$ from 0 to 3 depend on $t, r,
\theta, and~ \phi$ and the components of Ricci tensor on $r$ only.
I solve these set of ten partial differential equations for the
components of the Ricci collineation vector field. I consider
first Ricci collineation equations (9) and solve them
simultaneously to obtain the components of $\textbf{V}$ in terms
of arbitrary functions of the coordinates. Substituting these
values in other Ricci collineation equations I get conditions on
these arbitrary functions. Solving these conditions and checking
consistency with the Ricci collineation equations at every step
until these functions are determined explicitly and the final form
of $\textbf{V}$ involving arbitrary constants is obtained. While
finding these solutions,the constraints on the components of the
Ricci tensor are obtained. Solving these constraints will give the
metrics of the space time. I evaluate Ricci collineations only for
those cases which have non-degenerate Ricci tensor i.e , $det
(R_{\mu\nu} \neq 0)$. The degenerate cases for these spaces have
already been discussed in reference \cite{a23} . These equations
(9 - 13) have been solved simultaneously for the components of
collineation vector and get the solutions of these components in
the form in which they become a known functions of $\theta$ and
$\phi$ and unknown in these of $t$ and $r$, as given below
\begin{equation}\label{14}
V^{0} =
(\frac{-R_{2}}{R_{0}})f^{2}_{k}[(\int((\frac{-1}{f^{2}_{k}})(\int
f_{k}d\theta)d\theta)(A_{1}(t, r)_{,0}sin\phi - A_{2}(t,
r)_{,0}cos\phi)] + (\frac{-R_{2}}{R_{0}}) A_{3}(t, r)_{,0}(\int
f_{k}d\theta) + P(t, r)
\end{equation}
\begin{equation}\label{15}
V^{1} =
(\frac{-R_{2}}{R_{1}})f^{2}_{k}[\int(\frac{-1}{f^{2}_{k}})(\int
f_{k}d\theta)d\theta(A_{1}(t, r)_{,1}sin\phi - A_{2}(t,
r)_{,1}cos\phi)] + (\frac{-R_{2}}{R_{1}}) A_{3}(t, r)_{,1}(\int
f_{k}d\theta) + Q(t, r)
\end{equation}
\begin{equation}\label{16}
V^{2} = \int f_{k}d\theta ( A_{1}(t, r)sin\phi - A_{2}(t,
r)cos\phi) + C_{1}sin\phi - C_{2}cos\phi + A_{3}(t, r)f_{k}
\end{equation}
\begin{equation}\label{17}
V^{3} = (\int ((\frac{-1}{f^{2}_{k}})\int
f_{k}d\theta)d\theta)[A_{1}(t, r)cos\phi + A_{2}(t, r)sin\phi] +
(C_{1}cos\phi + C_{2}sin\phi) \int (\frac{-1}{f^{2}_{k}})d\theta +
C_{3}
\end{equation}
where $C_{1}$, $C_{2}$ and $C_{3}$ are arbitrary constants. Here
the partial derivatives with respect to $0$ and $1$, indicate the
derivatives with respect to $t$ and $r$ coordinates, respectively.
Replacing the values of $V^{\mu}$ with $\mu$ from 0 to 3 in Ricci
collineation equations (9-13) show that $(C_{01)}$, $(C_{03})$,
$(C_{12})$, $(C_{13})$, and $(C_{23})$ are satisfied identically,
whereas $(C_{00})$, $(C_{02})$, $(C_{11})$, $(C_{22})$, and
$(C_{33})$ are satisfied subject to the following differential
constraints on $A_{i}(t, r)$, $P(t, r)$ and $Q(t, r)$, with $C_{4}
= 0$.
\begin{equation}\label{18}
R^{\prime}_{0}A^{\prime}_{i}(t, r) + 2R_{1}A_{i}(t, r)_{,00} =
0,~~~~~~ i = 1,~ 2,~ 3
\end{equation}
\begin{equation}\label{19}
[\sqrt{\frac{R_{2}}{R_{0}}}A_{i}(t, r)_{,0}]^{\prime} = 0
\end{equation}
\begin{equation}\label{20}
[\frac{R_{2}}{\sqrt{R_{1}}}A^{\prime}_{i}(t, r)]^{\prime} = 0
\end{equation}
\begin{equation}\label{21}
\frac{R^{\prime}_{2}}{2R_{1}}A^{\prime}_{i}(t, r) - A_{i}(t, r) =
0
\end{equation}
\begin{equation}\label{22}
P(t, r)_{,0} + \frac{R^{\prime}_{0}}{2R_{0}}Q(t, r) = 0
\end{equation}
\begin{equation}\label{23}
P^{\prime}(t, r) + \frac{R_{1}}{R_{0}}Q(t, r)_{,0} = 0
\end{equation}
\begin{equation}\label{24}
(R_{2}\sqrt{R_{1}}Q(t, r))^{\prime} = 0
\end{equation}
\begin{equation}\label{25}
R^{\prime}_{2}Q(t, r) = 0
\end{equation}
where $i = 1,....3$. In this article, The symmetry properties of
static space times with maximal symmetric transverse spaces are
investigated by considering Ricci collineation. Ricci collineation
equations are solved for the components of collineation vector
field by writing them in a form in which they become known
functions of $\theta$ and $\phi$ and unknown in these of $t$ and
$r$. Substituting these components into each of the collineation
equations imply that some of these equations are identically
satisfied, whereas others are not and get replaced by a set of
constraint equations to be solved for classifying collineations.
These collineations are explicitly derived in sec. 2. The
solutions of the constraints on the components of Ricci tensor and
metrics corresponding to these constraints are provided in sec. 3.
Finally, the results are summarized in sec. 4
\section{Classification}\label{2}
The constraint equations (18 - 25) are used to classify the space
times under consideration. The classification of the space times
under consideration is started by using equation (25). This
equation can be satisfied for three cases;
$$(I):~~~~~~~ R^{\prime}_{2} = 0 ~~~ and~~~ Q(t, r) \neq 0 $$
$$(II):~~~~~~ R^{\prime}_{2} \neq 0 ~~~ and~~~ Q(t, r) = 0 $$
$$(III):~~~~~ R^{\prime}_{2} = 0 ~~~ and~~~ Q(t, r) = 0 $$
\subsection{ \textbf{CASE (1)}~~~ [when $R^{\prime}_{2} = 0 ~~~implies
~~~R_{2} = \gamma \neq 0~~ and~~ Q(t, r) \neq 0$]} In this case,
it follows from equation (21) that $A_{i}(t, r) = 0$. Replacing
$A_{i}(t, r) = 0$ into constraint equations(18-25), indicate that
the equations (18-21) are satisfied identically while the others
get replace into the constraints on $P(t, r)$ and $Q(t, r)$.
Integrating equations (24) yields
\begin{equation}\label{26}
Q(t, r) = f(t)/\gamma\sqrt{R_{1}}
\end{equation}
where $f(t)$ is a function of integration with respect to $r$.
With this value of $Q(t, r)$,  I consider two more possibilities
from equation (22), namely,\\ \textbf{(a)} ~~~$R^{\prime}_{0} = 0$
~~ and ~~ \textbf{(b)} ~~~~$R^{\prime}_{0} \neq 0$\\
\textbf{Case(1a)} : ~~In this case, $R_{0} = \alpha$, where
$\alpha$ is a non-zero constant. Also it follows from equation
(22) that $P(t, r)_{,0} = 0$ implies $P \equiv P(r)$. Using these
values in equation (23), yields
\begin{equation}\label{27}
P(r) = (\frac{-\dot{f(t)}}{\alpha\gamma}) \int \sqrt{R_{1}} dr +
g(t)
\end{equation}
where $g(t)$ is another function of integration. Since $\dot{P(r)}
= 0$ implies $\ddot{f(t)} = 0 = \dot{g(t)}$, which lead to
$$f(t) = C_{4}t + C_{5}~~and~~ g(t) = C_{0}$$. Putting these
results into the collineation equations (14-17), leads to six
collineations
$$V^{0} = (\frac{-C_{4}}{\alpha\gamma}) \int \sqrt{R_{1}} dr +
C_{0}$$
$$V^{1} = \frac{1}{\gamma\sqrt{R_{1}}}(C_{4}t + C_{5})$$
$$V^{2} = C_{1}sin\phi - C_{2}cos\phi$$
$$V^{3} = \int (\frac{-1}{f^{2}_{k}}) d\theta (C_{1}cos\phi + C_{2}sin\phi) +
C_{3}$$\\\textbf{Case~(1b)}:~~~~~In this case, it follows from
equations (22), (23) and (26) that
\begin{equation}\label{28}
\gamma P(t, r)_{,0} =
(\frac{-R^{\prime}_{0}}{2R_{0}\sqrt{R_{1}}})f(t)
\end{equation}
\begin{equation}\label{29}
\gamma P(t,r)_{,1} = (\frac{-\sqrt{R_{1}}}{R_{0}})f(t)_{,0}
\end{equation}
Differentiating equations (28) and (29) with respect to $r$ and
$t$ respectively, and comparing them, one reaches
\begin{equation}\label{30}
f(t)_{,00} - (\frac{R_{0}}{\sqrt{R_{1}}})
(\frac{R^{\prime}_{0}}{2R_{0}\sqrt{R_{1}}})^{\prime} f(t) = 0
\end{equation}
The above equation suggests further two possibilities;\\
\textbf{(1bi)}~~$f(t) = 0$ and \textbf{(1bii)}~~ $f(t) \neq 0$\\
\textbf{Case~(1bi)}:~~~~~In this case, it follows from equation
(26), (22) and (23) that
$$Q(t, r) = 0 ~~~ and ~~~ P(t, r) = C_{0}$$
Using these results along with the fact that $A_{i}(t, r) = 0$,
imply that the space times under consideration admit four
collineations, given below
$$V^{0} = C_{0}$$
$$V^{1} = 0$$
$$V^{2} = C_{1}sin\phi - C_{2}cos\phi$$
$$V^{3} = (C_{1}cos\phi + C_{2}sin\phi) \int (\frac{-1}{f^{2}_{k}}) d\theta +
C_{3}$$\\\textbf{Case~(1bii)}:~~~~ Re-writing equation (30) in the
form
$$f(t)_{,00} / f(t) - (\frac{R_{0}}{\sqrt{R_{1}}})
(\frac{R^{\prime}_{0}}{2R_{0}\sqrt{R_{1}}})^{\prime}  = 0$$ In
this equation $f(t)$ is a function of $t$ only, whereas
$R_{0}~~and~~R_{1}$ are functions of $r$ only. Hence this equation
can be satisfied if and only if
\begin{equation}\label{31}
f(t)_{,00} / f(t) = (\frac{R_{0}}{\sqrt{R_{1}}})
(\frac{R^{\prime}_{0}}{2R_{0}\sqrt{R_{1}}})^{\prime}  = \lambda
\end{equation}
where $\lambda$ is a separation constant. There are further three
possibilities for the values of $\lambda$ \\\textbf{(I):~ [when ~~
$\lambda = 0$ ]}:~~~Equation (31) leads to
\begin{equation}\label{32}
f(t) = C_{4}t + C_{5}
\end{equation}
\begin{equation}\label{33}
R_{0} = \alpha_{2} exp (2\alpha_{1} \int \sqrt{R_{1}} dr)
\end{equation}
where $\alpha_{1}$ and $\alpha_{2}$ are constants of integration
with $\alpha_{2} \neq =$. Using equations (32) and (33), (28) and
(29) lead to two equations in $P(t, r)_{,0}$ and $P(t, r)_{,1}$
which on integration with respect to $t$ and $r$ give two values
of $P(t, r)$ along with two functions of integration depending on
$r$ and $t$, respectively. Comparing these values of $P(t, r)$ and
requiring consistency fixes these functions to give
\begin{equation}\label{34}
P(t, r) = (\frac{-\alpha_{1}}{\gamma}) (\frac{C_{4}t}{2} + C_{5})t
- (\frac{C_{4}}{\gamma \alpha_{2}}) \int [(\sqrt{R_{1}})(exp
(-2\alpha_{1} \int \sqrt{R_{1}})dr)]dr + C_{0}
\end{equation}
In this case, it follows from equation (34) along with the
previous results that there are six Ricci collineations, given by

$$V^{0} = (\frac{-\alpha_{1}}{\gamma}) (\frac{C_{4}t}{2} + C_{5})t
- (\frac{C_{4}}{\gamma \alpha_{2}}) \int [(\sqrt{R_{1}})(exp
(-2\alpha_{1} \int \sqrt{R_{1}})dr)]dr + C_{0}$$

$$V^{1} = ( \frac{1}{\gamma \sqrt{R_{1}}}) (C_{4}t +
C_{5})$$
$$V^{2} = C_{1}sin\phi - C_{2}cos\phi$$
$$V^{3} = \int (\frac{-1}{f^{2}_{k}})d\theta (C_{1}cos\phi +
C_{2}sin\phi) + C_{3}$$
\textbf{(II):~~~ [when ~$\lambda < 0
~~or~~ \lambda
> 0$]} In these cases,from equation (31) the space times admit
collineations, given below
\begin{equation}\label{35}
V^{0} = (\frac{-R^{\prime}_{0}}{2\gamma R_{0}\sqrt{\lambda
R_{1}}}) ( C_{4}f^{1}_{\lambda} - C_{5}f^{2}_{\lambda}) + C_{0} /
\gamma
\end{equation}
\begin{equation}\label{36}
V^{1} = ( \frac{1}{\gamma \sqrt{R_{1}}}) (C_{4}f^{2}_{\lambda} +
C_{5}f^{1}_{\lambda})
\end{equation}
\begin{equation}\label{37}
V^{2} = C_{1}sin\phi - C_{2}cos\phi
\end{equation}
\begin{equation}\label{38}
V^{3} = \int (\frac{-1}{f^{2}_{k}})d\theta (C_{1}cos\phi +
C_{2}sin\phi) + C_{3}
\end{equation}
where the functions $f_{\lambda}^{1}$  are defined by
$$f^{1}_{\lambda} = sin\sqrt{\lambda}t~~ when ~~\lambda < 0$$~ and~ $$f_{\lambda}^{1} =
sinh\sqrt{\lambda}t ~~ when ~\lambda > 0$$. Similarly the values
of the function $f_{\lambda}^{2}$ are given by

$$f_{\lambda}^{2} = cos\sqrt{\lambda}t~~ when ~~\lambda < 0$$ and
$$f_{\lambda}^{2} = cosh\sqrt{\lambda}t~~ when ~\lambda > 0$$.
\subsection{\textbf{CASE (2):}, ~~~~[ When ~~~ $R^{\prime}_{2} \neq 0$~~and~~$Q(t, r) =0 $]}
In this case $R_{2}^{\prime} \neq 0$ implies through equation
(2.41) that $Q(t, r) = 0$, however the equation (2.40) satisfied
identically. Equations (2.38) and (2.49) yield $P(t, r) = C_{0}$
and solving equations (2.36) and (2.37) one gets
\begin{equation}\label{39}
A_{i}(t, r) = (\frac{R^{\prime}_{2}}{2R_{2}\sqrt{R_{1}}}) f_{i}(t)
\end{equation}
where $i = 1,~ 2~ 3$. Using this value of $A_{i}$ in constraint
equation (2.37) one reaches
\begin{equation}\label{40}
[~~\frac{1}{R_{2}} - (\frac{1}{\sqrt{R_{1}}})
(\frac{R_{2}^{\prime}}{2R_{2}\sqrt{R_{1}}})^{\prime}~~]~~ f_{i}(t)
= 0
\end{equation}
This leads to two possibilities\\
\textbf{Case(2a)}~~~$[~~\frac{1}{R_{2}} - (\frac{1}{\sqrt{R_{1}}})
(\frac{R_{2}^{\prime}}{2R_{2}\sqrt{R_{1}}})^{\prime}~~] = 0$ \\
\textbf{Case (2b)}~~~ $ [~~\frac{1}{R_{2}} -
(\frac{1}{\sqrt{R_{1}}})
(\frac{R_{2}^{\prime}}{2R_{2}\sqrt{R_{1}}})^{\prime}~~] \neq 0$\\
From \textbf{Case (2b)}, one concludes  $f_{i}(t) = 0$ which in
turn leads that $A_{i}(t, r) = 0$ which is a case similar to
\textbf{Case (1)}.\\
\textbf{Case (2a):}~~~~ [ when ~~$\frac{1}{R_{2}} -
(\frac{1}{\sqrt{R_{1}}})
(\frac{R_{2}^{\prime}}{2R_{2}\sqrt{R_{1}}})^{\prime}~ = 0$~~]~~~
Equation (2.53) implies that $f_{i}(t)$ are arbitrary functions of
$t$ . Using equation (2.52) in equation (2.35) yields
\begin{equation}\label{41}
(\frac{R^{\prime}_{2}}{\sqrt{R_{0} R_{1} R_{2}}})^{\prime}
f_{i}(t)_{,0} = 0
\end{equation}
This suggests further two cases \\ \textbf{(Case 2ai)}~~~ $
(\frac{R^{\prime}_{2}}{\sqrt{R_{0} R_{1} R_{2}}})^{\prime} \neq 0$
\\ \textbf{Case(2aii)} ~~~$ (\frac{R^{\prime}_{2}}{\sqrt{R_{0} R_{1}
R_{2}}})^{\prime} = 0$ \\ Consider \textbf{Case (2ai)}, using
equation (41) we get $f_{i}(t)_{,0} = 0$ leads to
\begin{equation}\label{42}
f_{i}(t) = C_{i + 3}
\end{equation}
where $i$ runs from 1 to 3. Using equation (39) and (21) yields
\begin{equation}\label{43}
A^{\prime}_{i}(t, r) = (\frac{\sqrt{R_{1}}}{R_{2}}) f_{i}(t)
\end{equation}
Using equations (42), (39) and (43) yield
\begin{equation}\label{44}
A^{\prime}_{i}(t, r) = (\frac{R^{\prime}_{2}}{\sqrt{R_{1}R_{2}}})
C_{i + 3}
\end{equation}
This leads to $A_{i}(t, r)_{,0} = 0$ and
\begin{equation}\label{45}
A^{\prime}_{i}(t, r) = (\frac{\sqrt{R_{1}}}{R_{2}}) C_{i + 3}
\end{equation}
Using the solutions of the constraint equations, the components of
Ricci collineation vector field are given by
$$V^{0} = C_{0}$$
$$V^{1} = (\frac{-1}{\sqrt{R_{1}}}) f^{2}_{k} [\int ((\frac{-1}{f^{2}_{k}})(\int (f_{k}d\theta))) d\theta ( C_{4}sin\phi - C_{5}cos\phi)] - C_{6}(\frac{1}{\sqrt{R_{1}}}) \int f_{k} d\theta$$
$$V^{2} = (\frac{R^{\prime}_{2}}{2R_{2}\sqrt{R_{1}}}) \int
f_{k}d\theta (C_{1}sin\phi - C_{2}cos\phi) + C_{1}sin\phi -
C_{2}cos\phi + (\frac{R^{\prime}_{2}}{2R_{2}\sqrt{R_{1}}})f_{k}
C_{6}$$
$$V^{3} = (\int ((\frac{-1}{f^{2}_{k}})(\int (f_{k}d\theta)))
d\theta)(\frac{R^{\prime}_{2}}{2R_{2}\sqrt{R_{1}}})(C_{4}cos\phi +
C_{5}sin\phi) + \int (\frac{-1}{f^{2}_{k}}) d\theta (C_{1}cos\phi
+ C_{2}sin\phi) + C_{3}$$ In this case, I get seven Ricci
Collineations.\\
\textbf{Case (2aii):}~~~~ $(\frac{R^{\prime}_{2}}{\sqrt{R_{0}
R_{1} R_{2}}})^{\prime} = 0$\\ Equation (41) implies that
$f_{i}(t)$ are arbitrary of functions of $t$ while equations (19)
to (21) are satisfied identically, whereas equation (18) yields
\begin{equation}\label{46}
f_{i}(t)_{,00} / f_{i}(t) =(\frac{R^{\prime}_{0}}{R^{\prime}_{2}})
 = \delta
\end{equation}
where $\delta$ is a separation constant. It contains further three
possibilities depending on the value of $\delta$. Hence the values
of $f_{i}(t)$ for different values of $\delta$ can be written as
\begin{equation}\label{47}
f_{i}(t) = C_{i + 3}cos\sqrt{\delta}t + C_{i + 6}sin\sqrt{\delta}t
~~~~~~~~~~~~(\delta > 0)
\end{equation}
\begin{equation}\label{48}
f_{i}(t) = C_{i +3}t + C_{i + 6}
~~~~~~~~~~~~~~~~~~~~~~~~~~~~~(\delta = 0)
\end{equation}
\begin{equation}\label{49}
 f_{i}(t) = C_{i +3}cosh\sqrt{-\delta}t + C_{i +
6}sinh\sqrt{-\delta}t~~~~~~~(\delta < 0)
\end{equation}
Using equations (38), (39) and (40) along with equations (2.52)
and (2.56) implies that
\begin{equation}\label{50}
A_{i}(t, r) = (\frac{R^{\prime}_{2}}{2R_{2}\sqrt{R_{1}}})[C_{i +
3}cos\sqrt{\delta}t + C_{i + 6}sin\sqrt{\delta}t]
~~~~~~~~~~~~(\delta > 0)
\end{equation}
\begin{equation}\label{51}
A_{i}(t, r) = (\frac{R^{\prime}_{2}}{2R_{2}\sqrt{R_{1}}})[C_{i
+3}t + C_{i + 6}] ~~~~~~~~~~~~~~~~~~~~~~~~~~~~~(\delta = 0)
\end{equation}
\begin{equation}\label{52}
 A_{i}(t, r) = (\frac{R^{\prime}_{2}}{2R_{2}\sqrt{R_{1}}})[C_{i +3}cosh\sqrt{-\delta}t + C_{i +
6}sinh\sqrt{-\delta}t]~~~~~~~(\delta < 0)
\end{equation}
\begin{equation}\label{53}
A^{\prime}_{i}(t, r) = (\frac{\sqrt{R_{1}}{R_{2}}})[C_{i +
3}cos\sqrt{\delta}t + C_{i + 6}sin\sqrt{\delta}t]
~~~~~~~~~~~~(\delta > 0)
\end{equation}
\begin{equation}\label{54}
A^{\prime}_{i}(t, r) = (\frac{\sqrt{R_{1}}}{R_{2}})[C_{i +3}t +
C_{i + 6}] ~~~~~~~~~~~~~~~~~~~~~~~~~~~~~(\delta = 0)
\end{equation}
\begin{equation}\label{55}
A^{\prime}_{i}(t, r) = (\frac{\sqrt{R_{1}}}{R_{2}})[C_{i
+3}cosh\sqrt{-\delta}t + C_{i +
6}sinh\sqrt{-\delta}t]~~~~~~~(\delta < 0)
\end{equation}
and one may get $P(t, r) = C_{0}$ and $Q(t, r) = 0$. Substituting
these values in the Ricci collineation equations, the values of
the components of the vector field are

\begin{equation}\label{56}
V^{0} =
(\frac{-R_{2}}{R_{0}})f^{2}_{k}(\int((\frac{-1}{f^{2}_{k}})(\int
f_{k}d\theta)(A_{1}(t, r)_{,0}sin\phi - A_{2}(t, r)_{,0}cos\phi))
+ (\frac{-R_{2}}{R_{0}}) A_{3}(t, r)_{,0}(\int f_{k}d\theta) +
P(t, r)
\end{equation}
\begin{equation}\label{57}
V^{1} =
(\frac{-R_{2}}{R_{1}})f^{2}_{k}(\int((\frac{-1}{f^{2}_{k}})(\int
f_{k}d\theta)(B_{1}(t, r)_{,1}sin\phi - B_{2}(t, r)_{,1}cos\phi))
+ (\frac{-R_{2}}{R_{1}}) A_{3}(t, r)_{,1}(\int f_{k}d\theta) +
Q(t, r)
\end{equation}
\begin{equation}\label{58}
V^{2} = \int f_{k}d\theta ( A_{1}(t, r)sin\phi - A_{2}(t,
r)cos\phi) + C_{1}(t, r)sin\phi - C_{2}(t, r)cos\phi + A_{3}(t,
r)f_{k}
\end{equation}
\begin{equation}\label{59}
V^{3} = (\int ((\frac{-1}{f^{2}_{k}})\int
f_{k}d\theta)d\theta)(A_{1}(t, r)cos\phi + A_{2}(t, r)sin\phi) +
(C_{1}cos\phi + C_{2}sin\phi) \int (\frac{-1}{f^{2}_{k}})d\theta +
C_{3}
\end{equation}
where the values of $A_{i}(t, r)$ and $A^{\prime}_{i}(t, r)$ are
given above for different values of $\delta$ and the values of
$\dot{A}$ can be obtained by differentiating equations (41), (42)
and (43) with respect to $t$. In this case, there are ten
collineations with a finite lie algebra of dimension ten.
\subsection{\textbf{CASE (3):}~~~~~ [$R^{\prime}_{2} = 0 ~~~ and~~~ Q(t, r) = 0 $]}
It can easily be verified that this case turn out to be the as
case (1bii).
\section{Exact solutions of Einstein equations Admitting Non-trivial Ricci Collineations }\label{sec3}
In the previous section , static space times with maximal
symmetric transverse spaces are classified according to their
Ricci collineation vectors. I worked out Ricci collineations for
non-degenerate Ricci tensor and classifird the space time, whereas
degenerate case for these space times have already been worked out
in reference \cite{a23}. While working out collineations,  many
constraints on the components of Ricci tensor are obtained. In
this section I solve these constraints on Ricci tensor for
non-degenerate tensor and present the explicit form of some
metrics admitting non-trivial Ricci collineations. Equation (2.40)
gives six Ricci collineations along with the constraint $R_{2} =
\gamma \neq 0$ and $R^{\prime}_{0} / 2R_{2}\sqrt{R_{1}} = \delta$.
The simplest case could be $\delta = 0$ which implies $R_{0} =
\alpha$. Assuming $R_{2} = k = R_{0}$ and solving equations (5-7)
for these values, yields
\begin{equation}\label{60}
(\frac{A^{\prime\prime}}{2} - \frac{A^{\prime}B^{\prime}}{4} +
\frac{A^{\prime 2}}{4} + \frac{A^{\prime}}{r}) = k e^{B - A}
\end{equation}
where $k = 0, ~\pm 1$ and
\begin{equation}\label{61}
B^{\prime} - A^{\prime} = \frac{2}{r}
\end{equation}
Solving these equations simultaneously I get the following exact
solution
$$ds^{2} = r^{n}e^{\frac{k m r^{2}}{8}}(dt^{2} - m r^{2} dr^{2}) - r^{2}(d\theta^{2} + f_{k}^{2}d\phi^{2})$$
where m and n are non-zero arbitrary constants of integration.
These space times admit four killing vectors and six Ricci
collineations. One can consider $B(r) = 0$ so that the metric
under consideration takes the form
$$ds^{2} = e^{A}dt^{2} - dr^{2} - r^{2}(d\theta^{2} + f_{k}^{2}d\phi^{2})$$
along with the constraint $R^{\prime}_{2} = 0$ , I get the
following metric
$$ds^{2} = ar^{(b-2k)}dt^{2} - dr^{2} - r^{2}(d\theta^{2} + f_{k}^{2}d\phi^{2})$$
where a and b are constant of integrations with $a \neq 0$. This
space times admit six Ricci collineations . Again for the case $B
= 0$, considering $R_{2}^{\prime} = \alpha \neq 0$, I get the
metric for $k = 0$
$$ds^{2} = e^{-2\alpha r}dt^{2}- dr^{2} - r^{2}(d\theta^{2} + f_{k}^{2}d\phi^{2})$$
This metric admits four Ricci collineations. If one assumes $B =
0$ along with a constraint $R^{\prime}_{2} = 0$, I get the
following exact solutions
$$ds^{2} = r^{\alpha} dt^{2}- dr^{2} - r^{2}(d\theta^{2} + f_{k}^{2}d\phi^{2})$$
$$ds^{2} = (\frac{1}{\alpha r^{2} + 1}) dt^{2}- dr^{2} - r^{2}(d\theta^{2} + f_{k}^{2}d\phi^{2})$$
These both metrics admit four Ricci collineations. The constraint
equations corresponding to the case of ten Ricci collineations
true in the case of the metric, given by
$$ds^{2} = ({r^{2} / a^{2} - 1}) dt^{2} - (\frac{1}{r^{2} / a^{2} - 1})dr^{2} - r^{2}(d\theta^{2} + f_{k}^{2}d\phi^{2})$$
which admits ten Ricci collineations. Here I presented few special
cases, however one can solve these constraints equations on the
components of Ricci tensor in general and may find more new exact
solutions.

\section{Summary and Conclusion}\label{4}
In this article, static space times with maximal symmetric
transverse spaces are classified according to their Ricci
collineations.The Ricci collineation equations have been solved
for non-degenerate Ricci tensor $(det.(R_{\alpha} \neq 0))$. I
have obtained the explicit form of collineation vectors along with
constraints on the components of the Ricci tensor. Solving these
constraints on Ricci tensor, the explicit forms of some exact
solutions of the Einstein Field equations are obtained. In cases
(1a) and (1bii), six  collineations are obtained whereas the cases
(1bi), (2ai) and (2aii) amounted to four, seven and ten
collineations respectively. Hence it is concluded that the Lie
algebras of the collineations for non-degenerate is finite i.e of
dimensions ten, seven, six or four. It was found previously that
the algebra for these spaces \cite{a23} when the Ricci tensor is
degenerate, is not always finite. Note that the classification of
these spaces also cover the classification of static spherically
symmetric space times as a spacial case for $k = 1$.


\end{document}